\documentclass[final,5p,times,twocolumn]{myclass}

\usepackage{fullpage}
\usepackage{graphicx}
\usepackage{multirow}
\usepackage{amsmath}
\usepackage{amssymb}
\usepackage[colorlinks,hyperindex]{hyperref}
\usepackage{lineno}

\begin{document}

\begin{frontmatter}

\title{Random coincidence of $2\nu2\beta$ decay events as a background source in bolometric $0\nu2\beta$ decay experiments
}

\author[KIN,ORS]{D.M.~Chernyak}
\author[KIN]{F.A.~Danevich}
\author[ORS]{A.~Giuliani}
\author[ORS]{E.~Olivieri}
\author[ORS]{M.~Tenconi}  
\author[KIN]{V.I.~Tretyak}

\address[KIN]{Institute for Nuclear Research, MSP 03680 Kyiv, Ukraine}
\address[ORS]{Centre de Spectrom\'etrie Nucl\'eaire et de Spectrom\'etrie de Masse, 91405 Orsay, France}

\begin{abstract}
Two neutrino double $\beta$ decay can create irremovable background
even in high energy resolution detectors searching for
neutrinoless double $\beta$ decay due to random coincidence of
$2\nu2\beta$ events in case of poor time resolution. Possibilities
to suppress this background in cryogenic scintillating bolometers
are discussed. It is shown that the present bolometric detector technologies enable to control this form of background at the level required to explore the inverted hierarchy of the neutrino mass pattern, including the case of bolometers searching for the neutrinoless double $\beta$ decay of $^{100}$Mo, which is characterized by a relatively short two neutrino double $\beta$ decay half-life. 
\end{abstract}

\end{frontmatter}

\section{Introduction}

Neutrinoless double beta ($0\nu2\beta$) decay is a key process in
particle physics since it provides the only experimentally viable possibility to test the
Majorana nature of neutrino and the lepton number conservation, establishing in the meantime the absolute scale and the hierarchy of the neutrino masses~\cite{DBD-rev}.

One of the most important goal of the next generation double
$\beta$ decay experiments is to explore the inverted hierarchy of
the neutrino masses. In the inverted scheme the effective Majorana $\langle m_{\nu} \rangle$
neutrino mass is expected to be in the interval $\sim 0.02-0.05$ eV. In order to check this range, an experimental
sensitivity (in terms of half-life) for the most promising nuclei
should be at the level $T_{1/2}\sim10^{26}-10^{27}$ yr. This requires a detector
containing a large number of studied nuclei ($\sim 10^{27}-10^{28}$),
with high energy resolution (at most a few \% at the energy of the decay $Q_{2\beta}$), large
(ideally 100\%) detection efficiency, very low (ideally zero) radioactive background.

Besides HPGe detectors used to search for $0\nu2\beta$ decay of $^{76}$Ge \cite{Klap01,Aals02},
cryogenic bolometers~\cite{Fio-Nii,JOLT2} -- luminescent \cite{GM,giusang,Mil-scint,Pirr06,Taba10} or not~\cite{Cuoricino,CUORE,Ales11} -- are excellent candidates to realize large-scale high-sensitivity experiments involving different isotopes with high energy resolution (a
few keV) and detection efficiency (near $70\%-90\%$, depending on the
crystal composition and size). 

The CUORE cryogenic experiment \cite{CUORE,Ales11}, built on the successful Cuoricino and searching for $0\nu2\beta$ decay of
$^{130}$Te with the help of TeO$_2$ detectors, is by far the most advanced bolometric search and is under construction now, while several searches (among which LUCIFER \cite{LUCIFER} and AMoRE \cite{Lee10}), aiming to use different luminescent bolometers to search for double $\beta$ decay  of $^{82}$Se (ZnSe
\cite{LUCIFER,Arna11a}), $^{116}$Cd (CdWO$_4$ \cite{Giro09}), $^{100}$Mo
(CaMoO$_4$ \cite{Lee10} and ZnMoO$_4$ \cite{Giro10,JOLT1,Beem11,Beem12}), and $^{130}$Te (TeO$_2$~\cite{Cere}) are in the R\&D stage. 

However, a disadvantage of cryogenic bolometers is their poor time resolution. This can lead to
a background component at the energy $Q_{2\beta}$ due to random coincidences of lower energy signals, especially those due to the unavoidable two neutrino double $\beta$ decay ($2\nu2\beta$) events. 

The random coincidence of $2\nu2\beta$ events as a source of background in high sensitivity $0\nu2\beta$ experiments was considered and discussed for the first time in \cite{Beem11}. In this work, the contribution of random coincidences (rc) of $2\nu2\beta$ events to the counting rate in the energy region of the expected $0\nu2\beta$ peak is estimated. Methods to suppress the background are discussed.

\section{Random coincidence of $2\nu2\beta$ events}
\label{sec:random}

Energy spectra of $\beta$ particles emitted in $2\nu2\beta$ decay are related to
the 2-dimen\-sio\-nal distribution $\rho_{12}(t_{1},t_{2})$
(see f.e.~\cite{DBD-tab} and refs. therein)
\begin{equation}
\label{eq:shape}
\rho_{12}(t_{1},t_{2})=e_1p_1F(t_1,Z) \cdot e_2p_2F(t_2,Z) \cdot (t_0-t_1-t_2)^5,
\end{equation}
where $t_i$ is the kinetic energy of the $i$-th electron
(all energies here are in units of the electron mass $m_0c^2$),
$t_{0}$ is the energy available in the $2\beta$ process,
$p_i$ is the momentum of the $i$-th electron $p_i=\sqrt{t_i(t_i+2)}$
(in units of $m_0c$), and $e_i=t_i+1$.
The Fermi function $F(t,Z)$, which takes into account the influence
of the electric field of the nucleus on the emitted electrons, is defined as
\begin{equation}
F(t,Z)=const \cdot p^{2s-2}exp(\pi\eta)\mid\Gamma(s+i\eta)\mid^2,
\end{equation}
where $s=\sqrt{1-(\alpha Z)^2}$, $\eta=\alpha Ze/p$, $\alpha=1/137.036$,
$Z$ is the atomic number of the daughter nucleus ($Z>0$ for $2\beta^-$
and $Z<0$ for $2\beta^+$ decay), and $\Gamma$ the gamma function.

The distribution $\rho(t)$ for the sum of electron energies $t=t_1+t_2$
is obtained by integration
\begin{equation}
\label{eq:integr}
\rho(t)=\int_0^t \rho_{12}(t-t_2,t_2)\,dt_2.
\end{equation}

\begin{figure}[h]
\centering
\includegraphics[width=0.4\textwidth]{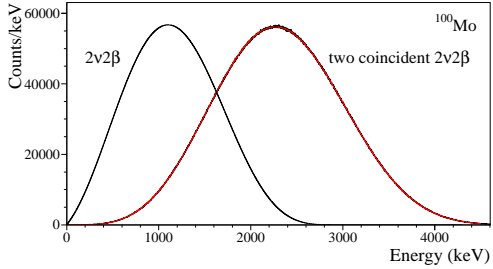}
\caption{(Color
online) Distribution for the sum of energies of two electrons
emitted in $2\nu2\beta$ decay of $^{100}$Mo and energy spectrum of
$10^8$ two randomly coincident $2\nu2\beta$ events for $^{100}$Mo
obtained by Monte Carlo sampling. The approximation of the random coincidence spectrum by the expression~(\ref{eq:compact}) is shown by the solid(red) line.}
\label{fig:distr}
\end{figure}

This distribution is shown for $^{100}$Mo in Fig.~\ref{fig:distr}.

The Primakoff-Rosen (PR) approximation for the Fermi function
$F(t,Z) \sim e/p$ \cite{Pri59}, which is adequate for $Z>0$,
allows to simplify Eq.~(\ref{eq:shape}) to the expression
\begin{equation}
\rho_{12}^{PR}(t_{1},t_{2})=(t_1+1)^2(t_2+1)^2(t_0-t_1-t_2)^5
\end{equation}
and to obtain the formula for $\rho(t)$ analytically:
\begin{equation}
\label{eq:analyt}
\rho^{PR}(t)=t(t_0-t)^5(t^4+10t^3+40t^2+60t+30).
\end{equation}

The energy distribution for two randomly coincident $2\nu2\beta$ decays
$\rho_{rc}(t)$ can be obtained by numerical convolution
\begin{equation}
\label{eq:conv}
\rho_{rc}(t)=\int_0^t \rho(t-x)\rho(x)\,dx,
\end{equation}
or with a Monte Carlo method by sampling energy releases in two independent
$2\nu2\beta$ events in accordance with the distribution~(\ref{eq:integr})
and adding them.
The energy spectrum obtained by sampling $10^8$ coincident
$2\nu2\beta$ events for $^{100}$Mo is shown in Fig.~\ref{fig:distr}. 
(We assume here an ideal energy resolution of the detector.) This distribution can be approximated by the following compact expression,\footnote{One can expect a polynomial of 21-st degree because of the formulae~(\ref{eq:analyt}) and (\ref{eq:conv}).} similar to that reported in Eq.~(\ref{eq:analyt}):
\begin{equation}
\label{eq:compact}
\rho_{rc}(t) = t^3 (2t_0-t)^{10} \sum_{i=0}^8 a_it^i.
\end{equation}

However, the coefficients $a_i$ are different for different
isotopes; for $^{82}$Se, $^{100}$Mo, $^{116}$Cd, $^{130}$Te
(which are the nearest aims of the bolometric $2\beta$ experiments),
they are given in Table~\ref{tab:coeff}. The approximation for $^{100}$Mo is shown in
Fig.~\ref{fig:distr}.

\begin{table}[h!]
\caption{Coefficients $a_i$ in the energy distribution~(\ref{eq:compact}) for two randomly
coincident $2\nu2\beta$ events for $^{82}$Se, $^{100}$Mo, $^{116}$Cd, $^{130}$Te.}
\label{tab:coeff}
\begin{center}
\begin{tabular}{lllll}
\hline\noalign{\smallskip}
~     & \multicolumn{4}{c}{Isotope} \\
$a_i$ & $^{82}$Se & $^{100}$Mo & $^{116}$Cd & $^{130}$Te \\
\noalign{\smallskip}\hline\noalign{\smallskip}
$a_0$ &  ~3446.59 &  ~5827.48  &  ~20093.8  &  ~15145.6  \\
$a_1$ & --7746.37 & --14399.7  & --1318.65  &  ~5554.78  \\
$a_2$ &  ~22574.7 &  ~34128.1  &  ~69134.6  &  ~39930.0  \\
$a_3$ & --16189.1 & --23815.5  & --31971.3  & --13338.5  \\
$a_4$ &  ~8467.01 &  ~11271.7  &  ~17976.8  &  ~7689.45  \\
$a_5$ & --2156.83 & --2711.31  & --3486.40  & --813.887  \\
$a_6$ &  ~337.172 &  ~390.396  &  ~406.762  & --80.3126  \\
$a_7$ & --28.9146 & --30.8774  & --30.5846  &  ~19.1035  \\
$a_8$ &  ~1.      &  ~1.       &  ~1.       & --1.       \\
\noalign{\smallskip}\hline
\end{tabular}
\end{center}
\end{table}

The random coincidence counting rate $I_{rc}$ in a chosen energy
interval $\Delta E$ is determined by the time resolution of the detector
$\tau$ and the counting rate for single $2\nu2\beta$ events $I_0$:
\begin{equation}\label{eq:irc}
I_{rc} = \tau \cdot I_0^2 \cdot \varepsilon =
\tau \cdot \left( \frac{\ln 2~N}{T^{2\nu2\beta}_{1/2}} \right)^2 \cdot \varepsilon,
\end{equation}
where $N$ is the number of $2\beta$ decaying nuclei under investigation,
and $\varepsilon$ is the probability of registration of events in the $\Delta E$ interval. In Eq.~(\ref{eq:irc}) and in the following, we assume that if two events occur in the detector within a temporal distance lower than the time resolution $\tau$ they give rise to a single signal with an amplitude equal to the sum of the amplitudes expected for the two separated signals.
The calculated probabilities at the energy $Q_{2\beta}$ of the $0\nu2\beta$ decay 
for $\Delta E = 1$ keV interval are equal to
$\varepsilon=3.5\times10^{-4}$ for $^{82}$Se and $\varepsilon=3.3\times10^{-4}$ for
$^{100}$Mo, $^{116}$Cd, $^{130}$Te.

Counting rates of detectors with 100 cm$^3$ volume  (typical for large mass bolometers)
at the energy of $0\nu2\beta$ decay for different $2\beta$
candidates and compounds are presented in Table~\ref{tab:contr}. We assume 100\%
isotopical enrichment for $^{82}$Se, $^{100}$Mo, $^{116}$Cd,
and natural abundance (34.08\%) for $^{130}$Te.
The time resolution of a detector is assumed as $\tau=1$ ms. The reported rates scale linearly with the time resolution.

Background $B_{rc}$ caused by random coincidences of $2\nu2\beta$ events
has the following dependence on the
energy resolution $R$,
the volume of the detector $V$,
the abundance or enrichment $\delta$ of the candidate nuclei contained in the detector:
\begin{equation}\label{eq:brc}
B_{rc} \sim \tau \cdot R \cdot (T_{1/2}^{2\nu2\beta})^{-2} \cdot V^2 \cdot \delta^{2}.
\end{equation}
One can conclude from Table~\ref{tab:contr} that the most important random coincidence background
is for $^{100}$Mo due to the relatively short $2\nu2\beta$ half-life.
However, a non-negligible contribution is expected also for other isotopes in case
of big volume and poor time resolution of the single detectors.

\begin{table*}[htb]
\caption{Counting rate of two randomly coincident $2\nu2\beta$ events
in cryogenic Zn$^{82}$Se, $^{40}$Ca$^{100}$MoO$_4$,
Zn$^{100}$MoO$_4$, $^{116}$CdWO$_4$ and TeO$_2$ detectors of 100 cm$^3$
volume. Enrichment of $^{82}$Se, $^{100}$Mo and $^{116}$Cd is assumed 100\%,
while for Te the natural isotopic abundance (34.08\%) is taken.
$C$ is the mass concentration of the isotope of interest,
$\rho$ is the density of the material (g/cm$^3$),
$N$ is the number of $2\beta$ candidate nuclei in one detector,
and $B_{rc}$ is the counting rate at $Q_{2\beta}$
(counts/(keV$\cdot$kg$\cdot$yr)) under assumption of 1 ms time
resolution of the detectors.}
\label{tab:contr}
\begin{center}
\begin{tabular}{llllll}
\hline\noalign{\smallskip}
Isotope    & $T_{1/2}^{2\nu2\beta}$ (yr)~\cite{Bara10} & Detector ($\rho$)               & $C$    & $N$                   & $B_{rc}$ \\
\noalign{\smallskip}\hline\noalign{\smallskip}
$^{82}$Se  & 9.2$\times$$10^{19}$        & Zn$^{82}$Se (5.65)              & 55.6\% & 2.31$\times$$10^{24}$ & 5.9$\times$$10^{-6}$ \\
$^{100}$Mo & 7.1$\times$$10^{18}$        & $^{40}$Ca$^{100}$MoO$_4$ (4.35) & 49.0\% & 1.28$\times$$10^{24}$ & 3.8$\times$$10^{-4}$ \\
~          & ~                           & Zn$^{100}$MoO$_4$ (4.3)         & 43.6\% & 1.13$\times$$10^{24}$ & 2.9$\times$$10^{-4}$ \\
$^{116}$Cd & 2.8$\times$$10^{19}$        & $^{116}$CdWO$_4$ (8.0)          & 31.9\% & 1.32$\times$$10^{24}$ & 1.4$\times$$10^{-5}$ \\
$^{130}$Te & 6.8$\times$$10^{20}$        & TeO$_2$ (5.9)                   & 27.2\% & 0.76$\times$$10^{24}$ & 1.1$\times$$10^{-8}$ \\
\noalign{\smallskip}\hline
\end{tabular}
\end{center}
\end{table*}

Obviously any other source of background with energy high
enough can contribute due to random coincidences. For example, the presence of $^{234m}$Pa (belonging to the series of $^{238}$U) with a relatively high activity of 1 mBq/kg in 100 cm$^3$ ZnMoO$_4$ crystal will result in additional background due to random coincidence with $2\nu2\beta$ events of
$3.8\times10^{-5}$ counts/(keV$\cdot$kg$\cdot$y) at $^{100}$Mo
$Q_{2\beta}$ energy. This contribution is however much less important than that due to the $2\nu2\beta$ decay alone. In addition, our simulations show that coincidence of $2\nu2\beta$ signals with low energy events is not problematic, because of the quite steep shape of $2\nu2\beta$ spectrum near $Q_{2\beta}$. Low energy signals due to spurious sources, like microphonic noise, which can in principle contribute with a high rate, are in general easily rejected in bolometers thanks to pulse shape discrimination.
In conclusion, while a generic source can be reduced by careful shielding, purification of materials, improvement of the noise figure and anti-coincidence technique, the random coincidence of $2\nu2\beta$ events is a background hard to suppress, as it is related to the presence itself of the isotope under investigation. The possibilities to decrease this type of background at an acceptable level are discussed in the next Section.

\section{Pile-up rejection in scintillating bolometers}
\label{sec:bolo}

A cryogenic bolometer~\cite{bolo} consists of an energy absorber (a single diamagnetic dielectric crystal in $0\nu2\beta$ applications) thermally linked to a temperature sensor, that in some cases may be sensitive to out of equilibrium phonons. The heat signal, collected at very low temperatures (typically $<20$~mK for large bolometers), consists of a temperature rise of the whole detector determined by a nuclear event.  

The majority of the most promising high $Q_{2\beta}$-value ($> 2.5$~MeV) candidates can be studied with the bolometric technique in the ``source=detector'' approach, joining high energy resolution and large efficiency~\cite{Fio-Nii,JOLT2}. Ultra-pure crystals up to $100-1000$~g can be grown with materials containing appealing candidates. Arrays of the single crystalline modules allow achieving total masses of the order of $100-1000$~kg~\cite{CUORE,Ales11,LUCIFER}, necessary to explore the inverted hierarchy region. 

An excellent choice for the bolometric material, performed in the Cuoricino and CUORE experiments~\cite{Cuoricino,CUORE}, consists of TeO$_2$ (tellurite) that has a very large (27\% in mass) natural content of the $0\nu2\beta$ candidate $^{130}$Te. In terms of background, the experience provided by TeO$_2$ searches~\cite{CUO-bkg} shows clearly that energy-degraded $\alpha$~particles, emitted by the material surfaces facing the detectors or by the detector surfaces themselves, are expected to be the dominant contribution in all high $Q_{2\beta}$-value candidates, for which the signal falls in a region practically free of $\gamma$~background. The $\alpha$ background component however can be made negligible using scintillating or in general luminescent (including Cherencov light~\cite{Taba10,Cere}) bolometers. In fact, since the $\alpha$ light yield is generally appreciably different from the $\beta$/$\gamma$ light yield at equal deposited energy, while the thermal response is substantially equivalent, the simultaneous detection of light and heat signals, and the comparison of the respective amplitudes, represents a powerful tool for $\alpha$/$\beta$ discrimination and therefore for $\alpha$ background rejection~\cite{GM,giusang,Mil-scint}. Scintillation photons are usually detected by a dedicated bolometer, in the form of a thin slab, opaque to the emitted light and equipped with its own temperature sensor. The light absorber, normally a Ge, Si or Si-coated Al$_2$O$_3$ slab, is placed close to a flat face of the main scintillating crystal. 

A wealth of prelimimary experimental results ~\cite{Pirr06,Arna11a,Giro09,Giro10,JOLT1,Beem11,Beem12} show that this method is effective and $\alpha$ rejection factors even much better than $99.9$\% can be achieved. Once that this discrimination capability of scintillating bolometer is taken into account, a detailed background analysis, based on reasonable assumptions on internal radioactive contamination, shows that a residual background level of $\sim 10^{-4}$ counts/(keV$\cdot$kg$\cdot$y) can be safely assumed~\cite{Beem11,Beem12}, opening the opportunity to explore the inverted hierarchy region of the neutrino mass pattern. However, the random coincidences of $2\nu2\beta$ events discussed in Section~\ref{sec:random} need to be kept under control. According to Eq.~(\ref{eq:irc}), the rate of rc-$2\nu2\beta$ events is proportional to the time
resolution of the detector. Therefore the time properties of the signal from a cryogenic detector play a crucial role in this form of background.

The large mass ($\sim 800$~g), high energy resolution ($\sim 3-4$~keV~FWHM at $2615$~keV) detectors developed for Cuoricino and CUORE~\cite{Cuoricino,CUORE} represent a sort of paradigma for $0\nu2\beta$ decay bolometers, inspiring also other proposed experiments and the related R\&D activities. As temperature sensors, Neutron Transmutation Doped (NTD) Ge elements are used, characterized by a high impedance (1-100 M$\Omega$) and a high sensitivity ($-d\log R / d \log T \sim 10$).  Other promising solutions for the thermal sensors have been used~\cite{Lee10,Flei09,TES,NbSi}, but for the moment only the energy resolution and the reliability provided by NTD-Ge-based bolometers seem to be compatible with a large-scale $0\nu2\beta$ experiment. In these devices, the NTD Ge thermistor is glued to the TeO$_2$ crystal by means of a two-component epoxy. Their time resolution is strictly related to the thermal-signal risetime. 

The temporal behaviour of the thermal pulse, and therefore its risetime, can be understood thanks to a thermal model for the whole detector, described elsewhere \cite{A9,A11,PED,SSB2}. The model predicts that large mass detectors with NTD-Ge readout have risetimes of the order of tens of milliseconds. This is confirmed experimentally in the Cuoricino detectors, which exhibited pulse risetimes of the order of $\sim 50$~ms~\cite{Cuoricino}. Similar values are expected for any cryogenic bolometers with a volume of the order of 100 cm$^3$ and based on NTD~Ge~thermistors. Faster risetimes could be observed if an important component of the energy reaches the thermistors in the form of athermal phonons, but this possibility depends critically on the nature of the main crystal and of the crystal-glue-thermistor interface. The most conservative approach consists in assuming the slow risetime evaluated and observed in TeO$_2$ bolometers.

The time resolution of the main crystal can be made substantially shorter than the risetime, taking advantage of the excellent signal-to-noise ratio expected at the $0\nu2\beta$ energy~\cite{vbt}, which is of the order of 2000 to 1 in TeO$_2$ crystals. (However, this high value is still to be proved for other bolometric materials relevant for $0\nu2\beta$ decay.) Even though the time resolution proved shorter by a factor 10 with respect to the present TeO$_2$ risetime values, and therefore around 5~ms, the background values reported in Table~\ref{tab:contr} should be multiplied by a factor 5, bringing them above $10^{-3}$~counts/(keV$\cdot$kg$\cdot$yr) for $^{100}$Mo. Background due to random coincidence of $2\nu2\beta$ events would be then dominant. The advantage of scintillating bolometers, which promise to keep the other sources around or below $10^{-4}$~counts/(keV$\cdot$kg$\cdot$yr) \cite{Beem11,Beem12}, would be substantially compromised.  

However, the simultaneous detection of light and heat which characterizes scintillating bolometers offers the possibility to control this problematic background source as well. In fact, a much faster risetime in the light-detector signal is expected, due to the much lower heat capacity of the energy absorber, which has now a mass of only a few grams at most. Following the bolometer thermal model~\cite{A9,A11,PED,SSB2}, the light-detector risetime can be reduced down to $\sim 1$~ms. Assuming a similar time resolution, the $2\nu2\beta$ background contribution is in the ranges shown in Table~\ref{tab:contr} thanks to the fast response provided by the light detector.

\begin{figure}[ht!]
    \centering
    \includegraphics[width=0.4\textwidth]{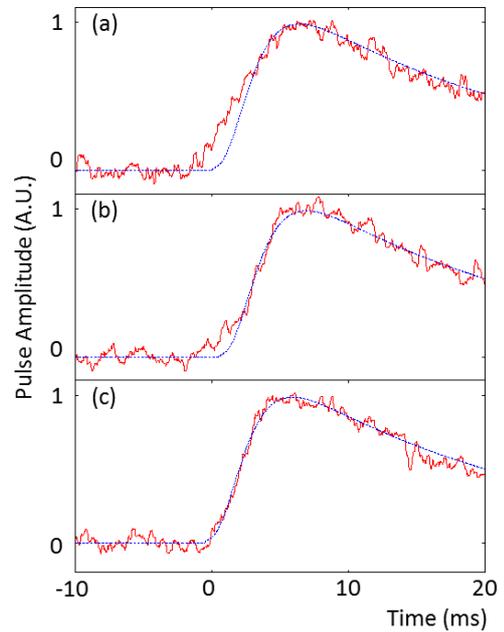}
    \caption{(Color online) Simulated piled-up pulses (solid/red lines) using real pulse shape and noise from a working light detector coupled to a ZnMoO$_4$ scintillating bolometer: (a) pile-up of two pulses shifted by 3 ms with equal amplitude; (b) pile-up of two pulses shifted by 3 ms with amplitude ratio equal to 4 (the smaller pulse occurs first); (c) a single pulse. The typical single-signal pulse shape, obtained by fitting an average pulse, is plotted as well (dashed/blue lines). In all cases, the signal-to-noise ratio is that expected for a $0\nu2\beta$ signal. The difference in shape between piled-up and single pulses is small, especially for unequal amplitudes, but appreciable.}
    \label{fig:pulse}  
\end{figure}

The above discussion is simplified: the capability to discriminate two close-in-time events cannot be reduced to a single parameter such as the detector time resolution $\tau$. In fact, it depends smoothly on the temporal distance between the two events and on the ratio between their two amplitudes as well. Furthermore, the pile-up discrimination capability is influenced by the signal-to-noise ratio in the light detector~\cite{vbt} and by the pulse shape and noise features. A complete analysis, whose details will be reported elsewhere in a dedicated publication, has been performed. In this letter, we present the main results of this investigation and the most important conclusions. 

An experiment based on modules of Zn$^{100}$MoO$_4$ crystals was considered. We have used pulse shapes and noise from a {\em real} light detector, of the same type as those described in~\cite{JOLT1,Beem11}, coupled to a ZnMoO$_4$ scintillating crystal. The pile-up phenomenon was studied by generating light pulses with the observed experimental shape on top of experimental noisy baselines. In particular, pair of pulses were generated with random time distances with a flat distribution up to 10~ms -- in fact, the interarrival time distribution is practically constant over the $[0,10]$~ms range (this is the relevant time interval since it allows to fully explore the most problematic pile-up case, i.e. that occurring on the pulse risetime which is of the order of 3~ms). In Fig.~\ref{fig:pulse}, two pile-up emblematic cases (both with 3~ms time separation) extracted from the performed simulation are shown, along with a single pulse as a reference.

As a first step, we defined a 90\% efficiency in accepting a pulse from the light detector as a potentially good $0\nu2\beta$ pulse using opportune signal filtering and three different pulse-shape indicators: (i) the risetime from 15\% to 90\% of the maximum amplitude; (ii) the $\chi^2$ evaluated using an average pulse as a standard shape function; (iii) the pulse shape parameter defined in~\cite{PSI} which also uses a standard pulse-shape function. The rejection efficiency of piled-up pulses was then tested. In each pulse pair, the amplitude of the first pulse $A_1$ was extracted by sampling the $2\nu2\beta$ distribution, while the amplitude of the second pulse $A_2$ was chosen as $Q_{2\beta}$($^{100}$Mo)$-A_1+\Delta E$, where $\Delta E$ is a random component in the interval $[-5,+5]$~keV.

The generated pulse amplitudes were chosen so as to fix the signal-to-noise ratio at the level expected for a $0\nu2\beta$ signal, i.e. of the order of 30, as shown in Fig.~\ref{fig:pulse}; in fact, the typical light energy collected by the light detectors in ZnMoO$_4$ scintillating bolometers realized so far is of the order of 1~keV for 1~MeV energy in the heat channel~\cite{Giro10,JOLT1,Beem11,Beem12}, while the typical RMS noise of the light detector can be conservatively taken as 100~eV, although values as low as 30~eV were observed~\cite{JOLT1}.

\begin{figure}[h]
\centering
\includegraphics[width=0.45\textwidth]{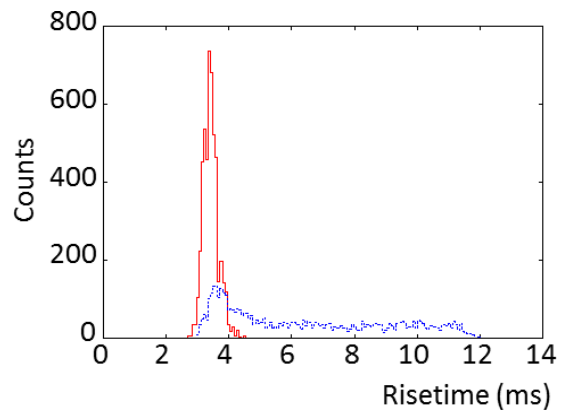}
\caption{(Color online) Risetime distribution for two populations of 5000 generated events each. The solid (red) line refers to single pulses; the dashed (blue) line is obtained with piled-up pulses separated by a time distance covering uniformly the interval $[0,10]$~ms, with amplitudes sampling the $2\nu2\beta$ spectrum and adding so as to fall in the region of the $0\nu2\beta$ expected peak.}
\label{fig:rtdistr}
\end{figure}

The piled-up pulses generated in the simulation were analyzed with the mentioned pulse-shape indicators. Using the risetime (after low-pass filtering), an excellent pile-up rejection efficiency was obtained. A comparison between the risetime distribution for genuine single pulses and piled-up pulses generated as described above is shown in Fig.~\ref{fig:rtdistr}. More quantitatively, the same procedure that retains 90\% of genuine single pulses rejects 80\%--90\% of piled-up pulses when their sum amplitude is in the region of $Q_{2\beta}$ and the difference between the arrival times of the two pulses covers uniformly the interval $[0,10]$~ms. For example, the analysis of the sample reported in Fig.~\ref{fig:rtdistr} excludes 83\% of piled-up pulses when accepting 90\% of good pulses. The other two indicators provide equivalent or even better results. However, we prefer here to consider conservatively the results obtained with the method of the risetime, as this parameter is an intrinsic property of each signal that does not require the comparison with a standard shape. This comparison in fact implies a delicate synchronization between the single pulse and the standard-shape pulse that will be discussed in the mentioned more complete work.

The results of the simulation show that the contribution to the background of the piled-up events is substantially equivalent to that obtained when assuming a time resolution $\tau$ of 1~ms in Eq.~\ref{eq:brc} (since 80\% -- 90\% of the pulses are rejected in the region of $0\nu2\beta$ decay inside a pile-up relevant range of 10~ms), and therefore confirming the evaluation for ZnMoO$_4$ reported in Table~\ref{tab:contr}. 

We can conclude therefore that light detectors at the present technological level are compatible with next-generation $0\nu2\beta$ decay experiments based on ZnMoO$_4$ crystals with background in the $10^{-4}$ counts/(keV$\cdot$kg$\cdot$y) scale, confirming that this class of experiments has the potential to explore the inverted hierarchy region of the neutrino mass pattern~\cite{Beem11,Beem12}. 

\section{Conclusions and prospects}

Random coincidence of $2\nu2\beta$ events is an
irremovable background source in a large scale $2\beta$
experiments using detectors with slow response time, such as large mass cryogenic bolometers with NTD~Ge readout.  

Advancement of time resolution of cryogenic detectors plays a key role to suppress the background. However, we have shown that the present technology is already compatible with searches at the sensitivity frontier. For further improvements, experimental efforts should be
concentrated on the time properties of the light signal, potentially much faster than the heat pulses of a scintillating bolometer. Achieving a time resolution below 0.1 ms could make the background totally negligible even for the difficult case of $^{100}$Mo. Such a performance could be obtained by using sensors sensitive to out-of-equilibrium phonons or intrinsically fast~\cite{Flei09,TES,NbSi}. 

A more direct way to decrease the pile-up effect is to reduce the volume of the main absorber (and increasing correspondingly the number of array elements), on which the random concindence rate depends quadratically, as shown in Eq.~(\ref{eq:brc}). Cryogenic detectors with space resolution could allow to reduce the background further.

\section{Acknowledgments}

The work of F.A.~Danevich and V.I.~Tretyak was supported in part
through the Project ``Kosmomikrofizyka-2'' (Astroparticle Physics)
of the National Academy of Sciences of Ukraine. The light detector results used for the pile-up simulation have been obtained within the project LUCIFER, funded by the European Research
Council under the EU Seventh Framework Programme (ERC grant agreement n. 247115). The background study in ZnMoO$_4$ scintillating bolometers is part of the program of ISOTTA, a project receiving funds from the ASPERA 2nd Common Call dedicated to R\&D activities.

\end{document}